\documentclass{elsart}

\usepackage{epsfig}
\usepackage{amssymb}
\usepackage{amsmath}
\usepackage{graphicx}
\usepackage{subfigure}
\usepackage{alltt}
\newcommand{\R}{\mathbb{R}}
\newcommand{\Z}{\mathbb{Z}}
\begin{document}

\begin{frontmatter}

\journal{Acta Cybernetica}

\title{Approximation of the Euclidean distance by chamfer distances}

\author[hajdua]{Andr\'as Hajdu\corauthref{cor}}
\corauth[cor]{Corresponding author.}
\ead{hajdu.andras@inf.unideb.hu}
\author[hajdul]{Lajos Hajdu}
\author[tijdeman]{Robert Tijdeman}

\address[hajdua]{Faculty of Informatics, University of Debrecen, H-4010
Debrecen, P.O.Box 12.}
\address[hajdul]{Number Theory Research Group of the Hungarian Academy of
Sciences, and Institute of Mathematics, University of Debrecen, H-4010
Debrecen, P.O.Box 12.}
\address[tijdeman]{Mathematical Institute, Leiden University, NL-2300 RA
Leiden, Postbus 9512.}

\begin{abstract}
Chamfer distances play an important role in the theory of distance
transforms. Though the determination of the exact Euclidean distance transform
is also a well investigated area, the classical chamfering method based upon
"small" neighborhoods still outperforms it e.g. in terms of computation time.
In this paper we determine the best possible maximum relative error
of chamfer distances under various boundary conditions. In each case some
best approximating sequences are explicitly given. Further, because of possible
practical interest, we give all best approximating sequences in case of small
(i.e. $5\times 5$ and $7\times 7$) neighborhoods.
\end{abstract}

\begin{keyword} Chamfering\sep Approximation of the Euclidean distance\sep
Distance transform\sep Digital image processing

\MSC 41A50 \sep 68U10

\end{keyword}
\end{frontmatter}

\section{Introduction}
\label{intro}

Suppose we measure distances between grid points of a two-dimensional
grid and we want to approximate the Euclidean distance by a distance
function which can be computed quickly, without calculating square
roots. We may then use the class of chamfer distances. They are obtained
by prescribing the lengths of the grid vectors in a so-called mask
$M_p :=\{(x,y)\in\Z^2:\max(|x|,|y|)\leq p\}$ (for some positive integer
$p$) such that the values at $(\pm x,\pm y)$ and $(\pm y,\pm x)$ are all
the same, and by defining the length function $W$ as follows: the length
$W(\vec{v})$ of any vector $\vec{v}\in\Z^2$ is defined as the minimal sum
of the lengths of those vectors from $M_p$, repetitions permitted, which
have sum $\vec{v}$.
The literature on chamfer distances is very rich. See Borgefors \cite{b1,b2,b3} for the basics, \cite{sc,sctij} for lists of $(2p+1) \times (2p+1)$ neighborhoods for $1\leq p \leq 10$,
and \cite{cuis} for an overview of applications. Further,  recently many related results have been obtained by several authors, concerning distance transforms and their explicit calculation using different kinds of neighborhoods in certain (mostly 3D) grids. For example, Strand, Nagy, Fouard and Borgefors \cite{snfb} gave a sequential algorithm for computing the distance map using distances based on neighborhood sequences in the 2D square grid, and 3D cubic and so-called FCC and BCC cubic grids, respectively. Similar results for other kinds of grids are also known, see e.g. \cite{ns12} ($n$D hexagonal grids), \cite{ns13} (diamond grid) and \cite{fsb} (general point grids) and the references given there.

Classical chamfer distances using $3\times 3$, $5\times 5$ and $7\times 7$
neighborhoods given by Borgefors \cite{b1,b2} are generated by the masks
$$
\begin{matrix}
\underline{4}&\underline{3}&\underline{4}\\
\underline{3}&0&\underline{3}\\
\underline{4}&\underline{3}&\underline{4}
\end{matrix}
\hskip1cm\mbox{,}\hskip1cm
\begin{matrix}
14&\underline{11}&10&\underline{11}&14\\
\underline{11}&\underline{7}&\underline{5}&\underline{7}&\underline{11}\\
10&\underline{5}&0&\underline{5}&10\\
\underline{11}&\underline{7}&\underline{5}&\underline{7}&\underline{11}\\
14&\underline{11}&10&\underline{11}&14
\end{matrix}
\hskip1cm\mbox{and}\hskip1cm
\begin{matrix}
51&\underline{43}&\underline{38}&36&\underline{38}&\underline{43}&51\\
\underline{43}&34&\underline{27}&24&\underline{27}&34&\underline{43}\\
\underline{38}&\underline{27}&\underline{17}&\underline{12}&\underline{17}&
\underline{27}&\underline{38}\\
36&24&\underline{12}&0&\underline{12}&24&36\\
\underline{38}&\underline{27}&\underline{17}&\underline{12}&\underline{17}&
\underline{27}&\underline{38}\\
\underline{43}&34&\underline{27}&24&\underline{27}&34&\underline{43}\\
51&\underline{43}&\underline{38}&36&\underline{38}&\underline{43}&51
\end{matrix}
$$
respectively (with the actual generator entries underlined).
For comparison with
the Euclidean distance the values of the neighborhoods have to be divided by
$3$, $5$ and $12$, respectively. The approximations to $\sqrt{2}\approx 1.41$
are therefore $4/3\approx 1.33$, $7/5=1.4$ and $17/12\approx 1.41$,
respectively. For alternative
neighborhood values see Verwer \cite{ve1,ve2}, Thiel \cite{th},
Coquin and Bolon \cite{cb}, Butt and Maragos \cite{bm} and Scholtus \cite{sc}.
More specifically, in \cite{cb} the minimization of the error between the Euclidean distance
and the local distance was considered over circular trajectories similarly to \cite{ve1,ve2}
rather than linear ones \cite{b2,th}. The approximation error can also be measured based on area
as it is done in \cite{bm} with calculating the difference between a disk of large size
obtained by chamfer metric and a Euclidean disk of the same radius.
The determination of the exact Euclidean distance transform is
also a well investigated area (see e.g. \cite{bailey,cuis,daniel,maurer,shih}),
but the classical $3\times 3$ chamfering method still outperforms it in terms
of computation time and simple extendability to other grids.

In this paper we determine chamfer distances best approximating the Euclidean
distance in a certain sense. In each neighborhood size some best approximating
sequences are explicitly given. Further, because of possible practical interest,
we give all best approximating sequences in case of small (i.e. $5\times 5$ and
$7\times 7$) neighborhoods.

Throughout the paper, as a measure for the quality of a length function
$W$ defined on $\Z^2$ we use the so-called maximum relative error
(m.r.error for short)
\[
E:=\limsup\limits_{|\vec{v}|\to\infty}
\left|{\frac{W(\vec{v})}{|\vec{v}|}}-1\right|
\]
where $|.|$ denotes the Euclidean length. The $M_1$-, $M_2$- and
$M_3$-neighborhoods given above yield rounded $E$-values $0.0572$, $0.0198$
and $0.0138$, respectively. Firstly we shall prove that the smallest possible constant
$E_p^B$ for the mask $M_p$ under the condition that $W(x,0)=|x|$ for $x\in\Z$
is given by
$$
E_p^B={\frac{p^2+2-p\sqrt{p^2+1}-2\sqrt{p^2+1-p\sqrt{p^2+1}}}{p^2}}
={\frac{1.5-\sqrt{2}}{p^2}}+O\left({\frac{1}{p^4}}\right).
$$
In particular, $E_1^B\approx 0.0551$, $E_2^B\approx 0.0187$ and
$E_3^B\approx 0.0089$. Comparing these values with the $E$-values given above, one can see that
the $E_p^B$-values yield approximately $4\%$, $6\%$ and $35\%$ improvement, respectively.
The $B$ refers to Borgefors who was the first to consider such neighborhoods.

Secondly we consider the case $D$ in which $W(\vec{v}) \geq |\vec{v}|$ for all $\vec{v}\in \Z^2$. (The $D$ refers to the fact that $W(\vec{v})$ dominates $|\vec{v}|$.)
The optimal m.r.error under this restriction equals
$$
E_p^D=\sqrt{(\sqrt{p^2+1}-p)^2+1}-1={\frac{1}{8p^2}}
+O\left({\frac{1}{p^4}}\right)
={\frac{0.125}{p^2}}+O\left( {\frac{1}{p^4}}\right).
$$
\noindent In particular, $E_1^D\approx 0.0824$, $E_2^D\approx 0.0275$ and $E_3^D\approx 0.0131$.

Thirdly we shall prove that the optimal $E$-value without any
restriction on the neighborhood defined on $M_p$ (i.e. dropping the condition $W(x,0)=|x|$ for $x\in\Z$) equals
$$
E_p^C={\frac{\sqrt{2p^2+2-2p\sqrt{p^2+1}}-1}
{\sqrt{2p^2+2-2p\sqrt{p^2+1}}+1}}
={\frac{1}{16p^2}}+O\left({\frac{1}{p^4}}\right).
$$
In particular, $E_1^C\approx 0.0396$, $E_2^C\approx 0.0136$ and $E_3^C\approx 0.0065$. In 1991, on using the symmetry in case $C$ the value of $E_p^C$ was computed by Verwer \cite{ve1,ve2} in terms of trigonometric functions. The $C$ refers to the word central. In 1998, because of geometric considerations, Butt and Maragos \cite{bm} chose to use the error function
\[
\limsup\limits_{|\vec{v}|\to\infty}
\left|{\frac{|\vec{v}|}{W(\vec{v})}}-1\right|
\]
which of course is small if and only if $E_p^C$ is small. In general it gives different error values, but the values for $E_p^C$ are equal
to the values obtained by the above error function
(cf. Scholtus \cite{sc}). We prove the correctness of the above $E_p^C$ values. In doing so, our motivation is twofold: on the one hand, by a simple reasoning we obtain these values immediately from the values of $E_p^D$, and on the other hand, our proofs are mathematically rigorous while the corresponding arguments of Verwer and Butt and Maragos contain some hidden assumptions. Namely, by certain plausible but not explicitly verified geometric arguments they restrict their attention and investigations to certain values of the neighborhoods in question, and they perform exact investigations only for these values.

We shall further study an auxiliary class of neighborhoods on $M_p$, viz. the class of neighborhoods satisfying $\mathcal{N}_c(\vec{v}) = \infty$ for all $\vec{v} = (x,y) \in M_p$ with either $x<p$ or $y<0$, $\mathcal{N}_c(\vec{v})=p$ for $\vec{v}=(p,0)$, and $\mathcal{N}_c(\vec{v}) = c|\vec{v}|$ for $\vec{v} = (p,k)$ with $0<k\leq p$. Here $c$ is a constant close to and at most equal to 1. Informally speaking, the use of such neighborhoods means that only such steps $(v_1,v_2)$ are allowed, where $v_1$ is a positive multiple of $p$ and $v_2$ is nonnegative. Further, beside ${\mathcal N}_c(p,0)=p$ the weights of the other such neighborhood vectors are their Euclidean lengths, multiplied by a factor $c\leq 1$. All the other vectors of the neighborhood are forbidden to use, thus they have weights $\infty$. For example, the weights for the neighborhood $\mathcal{N}_c$ with $p=2$ (i.e. for $M_2$) are given by
$$
\begin{matrix}
\infty&\infty&\infty&\infty&c\sqrt{8}\\
\infty&\infty&\infty&\infty&c\sqrt{5}\\
\infty&\infty&\infty&\infty&2\\
\infty&\infty&\infty&\infty&\infty\\
\infty&\infty&\infty&\infty&\infty
\end{matrix}
$$
where the origin is in the middle. We denote the maximum relative error for this class of neighborhoods by $\mathcal{E}^c_p$ where we restrict the
limsup to vectors $\vec{v}$ with finite lengths $W(\vec{v})$
(i.e. having coordiantes $(x,y)$ with $0\leq y\leq x$ and $p\mid x$). Our motivation for considering such neighborhoods is that it will turn out that (due to its special form) $\mathcal{N}_c$ is easier to handle, but yields the same m.r.error as the corresponding neighborhood $N_c$, in which $N_c(\pm p,0)=N_c(0,\pm p)=p$ and $N_c(x,y)=c\sqrt{x^2+y^2}$ otherwise ($(x,y)\in M_p$).

In Section 2 we introduce some notation and prove some preliminary results.
In Sections 3 and 4 we compute the values of $\mathcal{E}_p^B$ and $\mathcal{E}_p^D$
where $\mathcal{E}^B_p $ is the maximum relative error
$\mathcal{E}_p^c$ for optimal $c$ and $\mathcal{E}^D_p=\mathcal{E}_p^1$.
We give all sequences yielding minimal m.r.error in case of $5\times 5$ and
$7\times 7$ neighborhoods, as well.
In Section 4
we prove that $E_p^B = \mathcal{E}_p^B$ and $E_p^D = \mathcal{E}_p^D$
and further show that $E_p^C = E_p^D/(2+E_p^D)$ for all $p$.
Finally, we draw some conclusions in Section 5.

\section{Definitions and basic properties}
\label{defnot}

Let $N$ be a neighborhood defined on the mask $M_p$. Put
$M_p^*=M_p\setminus\{(0,0)\}$. We denote the value of $N$ at position $(n,k)$
by $w(n,k)$ for $(n,k)\in M_p$. Throughout the paper we assume that
$w(\pm n,\pm k)=w(\pm k,\pm n)>0$ for all
$(n,k)\in M_p^*$ and all possible sign choices. Hence it suffices to consider
the values $w(n,k)$ with $0\leq k\leq n\leq p$.

We can measure lengths of vectors and distances between points using neighborhood sequences.
Note that such sequences provide a flexible and very useful tool in handling several problems
in discrete geometry. For the basics and most important facts about such sequences,
see e.g. the papers \cite{dcc,yi,fhh,hht,nbuj} and the references given there. Here we only give
those notions which will be needed for our purposes.

Let $A=(N_i)_{i=1}^\infty$
be a sequence of neighborhoods defined on $M_p$ and $\vec{u},\vec{v}\in\Z^2$.
The sequence $\vec{u}=\vec{u}_0,\vec{u}_1,\hdots,\vec{u}_m=\vec{v}$ with
$\vec{u}_i-\vec{u}_{i-1}\in M_p$ is called an $A$-path from $\vec{u}$ to
$\vec{v}$. The $A$-length of the path is defined as
$\sum\limits_{i=1}^m w_i(\vec{u}_i-\vec{u}_{i-1})$. The distance
$W_A(\vec{v}-\vec{u})$ between $\vec{u}$ and $\vec{v}$, which is the $A$-length
of $\vec{v}-\vec{u}$, is defined as the minimal $A$-length taken over all
$A$-paths from $\vec{u}$ to $\vec{v}$. If the neighborhood sequence is fixed,
then we suppress the letter $A$ in the above notation.

If $N_i=N$ for all $i$, then the corresponding (constant) neighborhood
sequence is denoted by $A=\overline{N}$. We assume throughout the paper that
for such sequences $W(n,k)=w(n,k)$ holds for $(n,k)\in M_p$; if it would
not have been the case, then the function $w:=W|_{M_p^*}$ would have generated
$W$, too.

We call $W$ a metric if for all $\vec{u},\vec{v}\in\Z^2$
\begin{itemize}
\item $W(\vec{u}) < \infty$~~($W$ is finite),
\item $W(\vec{u})=0\Leftrightarrow \vec{u}=\vec{0}$ ($W$ is
positive definite),
\item $W(\vec{u})=W(-\vec{u})$ ($W$ is symmetric),
\item  $W(\vec{u}+\vec{v})\leq W(\vec{u})+W(\vec{v})$ ($W$ satisfies the triangle
inequality).
\end{itemize}

It follows from the above properties that $W(\vec{u})\geq 0$ for every $\vec{u} \in \Z^2$.
By our basic assumptions on $w$, every induced length function $W$ is
positive definite and symmetric. Furthermore, $W$ satisfies the triangle
inequality for $\vec{u}, \vec{v}$ with
$\vec{u}, \vec{v}, \vec{u}+\vec{v} \in M_p$ by definition.

The first lemma shows that in case of a constant neighborhood sequence
$W(\vec{v})/|\vec{v}|$ attains a minimal value which is reached already in $M_p^*$.

\begin{lem}
\label{three}
Let $N$ be a neighborhood defined on $M_p$ which induces the length
function $W$ on $\Z^2$. Then
$$
\liminf\limits_{|\vec{v}|\to\infty} {\frac{W(\vec{v})}{|\vec{v}|}}=
\min\limits_{\vec{v}\in M_p^*} {\frac{w(\vec{v})}{|\vec{v}|}}.
$$
\end{lem}

\begin{pf} Let $m=\min\limits_{\vec{v}\in M_p^*}
{\frac{w(\vec{v})}{|\vec{v}|}}={\frac{w(\vec{u})}{|\vec{u}|}}$
$(\vec{u}\in M_p^*)$. Then  for all $n$ we have
${\frac{W(n\vec{u})}{|n\vec{u}|}}=m$, so that
$\liminf\limits_{|\vec{v}|\to\infty} {\frac{W(\vec{v})}{|\vec{v}|}}\leq m$.
On the other hand, since ${\frac{w(\vec{v})}{|\vec{v}|}}\geq m$ for every
$\vec{v}\in M_p^*$, it follows from the definition of shortest path and the
triangle inequality for the Euclidean distance that
$$W(\vec{v})\geq \sum_i w(\vec{v}_i) = \sum_i \frac{w(\vec{v}_i)}{|\vec{v}_i|}\cdot |\vec{v}_i|
\geq m \sum_i |\vec{v}_i| \geq m|\vec{v}|$$
for every $\vec{v}\in\Z^2$ not equal to the origin.
Thus $\liminf\limits_{|\vec{v}|\to\infty}{\frac{W(\vec{v})}{|\vec{v}|}}\geq m$.
\qed
\end{pf}

The challenge is therefore to compute
$\limsup\limits_{|\vec{v}|\to\infty} {\frac{W(\vec{v})}{|\vec{v}|}}$.

\section{The maximum relative error for neighborhoods $\mathcal{N}_c$}
\label{secopt}

Let $c$ be some positive real number with $\frac{p}{\sqrt{p^2+1}} < c \leq 1$.
We shall study neighborhoods $\mathcal{N}_c$ on $M_p$ with
$\mathcal{N}_c(n,k)=\infty$ for which either $n<p$ or $k<0$, $\mathcal{N}_c(p,0)=p$ and $\mathcal{N}_c(p,k)=c\sqrt{p^2+k^2}$ for $0<k\leq p$. We are interested in the length function $\mathcal{W}_c$ induced by ${\mathcal A}_c:={\overline{{\mathcal N}_c}}$ for points in the set $\{(x,y) \in \Z^2 : p|x, 0 \leq y \leq x \}$. First we secure that under suitable conditions only two distinct steps occur in a shortest ${\mathcal A}_c$-path.

\begin{lem}
\label{twostep}
Let $\frac{p}{\sqrt{p^2+1}} <c \leq 1$.  Then a shortest ${\mathcal A}_c$-path from $(0,0)$ to $(mp,mr+k)$ with $m,r,k \in \Z, 0 \leq r<p, 0 \leq k <m$ consists only of steps $(p,r)$ and $(p,r+1)$.
\end{lem}

\begin{pf} Suppose a shortest path from $(0,0)$ to $(mp,mr+k)$ with
$m,r,k \in \Z, 0 \leq r<p, 0 \leq k <m$ contains two steps $(p,t)$ and $(p,u)$ with $t-2\geq u\geq 0$. Replace the two steps with steps $(p,t-1)$ and $(p,u+1)$, and write $L_1$ and $L_2$ for the length of the old and new paths, respectively.
Then we have
$$
L_1-L_2\geq c\sqrt{p^2+t^2}-c\sqrt{p^2+(t-1)^2}+
c\sqrt{p^2+u^2}-c\sqrt{p^2+(u+1)^2}=
$$
$$
=c(f_p(t)-f_p(u+1)),
$$
where
$$
f_p(x)=\sqrt{p^2+x^2} - \sqrt{p^2+(x-1)^2} \ \ \ (x\in \Z_{\geq 0}).
$$
A simple calculation yields that $f_p(x)$ is strictly monotone increasing in $x$, which shows that $L_1-L_2>0$. However, this contradicts the minimality of the length of the original path.

Hence a shortest path may contain steps $(p,t)$ and $(p,t+1)$ only, for some non-negative integer $t$. Since altogether we make $m$ steps, this immediately gives that $t=r$, and our statement follows.
\qed
\end{pf}

\begin{rem} The latter inequality is the most severe and explains why we restrict $c$ to
values greater than $\frac{p}{\sqrt{p^2+1}}.$
\end{rem}

\begin{cor}
\label{onestep}
Let $\frac{p}{\sqrt{p^2+1}} <c \leq 1$  Then a shortest ${\mathcal A}_c$-path from $(0,0)$ to $(mp,mr)$ with $0 \leq r \leq p$ consists of $m$ steps $(p,r)$.
\end{cor}

The next theorem gives the value of the approximation error for general $p$, in case of any neighborhood ${\mathcal{N}_c}$ on $M_p$.

\begin{thm}
\label{aaerror}
Let $p\geq 1$, $\frac{p}{\sqrt{p^2+1}}<c\leq 1$. Then the m.r.error of $\mathcal{A}_c$ to the Euclidean distance is given by
\[
\max(1-c,\sqrt{1+c^2+p^2+c^2p^2-2cp\sqrt{p^2+1}}-1).
\]
\end{thm}

\begin{pf}
As a general remark we mention that to perform our calculations, we used the
program package Maple \textsuperscript\textregistered
\footnote{Maple is a registered trademark of Waterloo Maple Inc.}.

Let $p$ be a positive integer, and fix $c$ with ${\frac{p}{\sqrt{p^2+1}}}<c\leq 1$. As previously, it is sufficient to consider the ${\mathcal A}_c$-length of points of the form $(mp,k)$ where $m$ is some positive integer and $k$ is an integer with $0\leq k\leq mp$. Write $k=mq+r$ with $0\leq q\leq p$ and $0\leq r<m$. The possible steps are $(p,0)$ of length $p$ and $(p,\pm i)$ of length $W_i:=c\sqrt{p^2+i^2}$ (for $|i|\leq p$). From Lemma \ref{twostep} and the inequalities $p=W_0<W_1<\hdots<W_p$ we see that a path of minimal length from $(0,0)$ to a point $(mp,mq+r)$ consists of $r$ steps $(p,q+1)$ and $m-r$ steps $(p,q)$. Hence
for the induced length function we get
\[
{\mathcal W}(mp,mq+r)=rW_{q+1}+(m-r)W_q.
\]
Put $t=r/m$, and recall that $W_0=p$ and $W_i=c\sqrt{p^2+i^2}$ for $i=1,\hdots,p$. Set
$$
H_0(t)={\frac{ct\sqrt{p^2+1}+p(1-t)}
{\sqrt{p^2+t^2}}}-1,
$$
and for $1\leq q\leq p$
$$
H_q(t)=c{\frac{t\sqrt{p^2+(q+1)^2}+(1-t)\sqrt{p^2+q^2}}
{\sqrt{p^2+(q+t)^2}}}-1,
$$
and let
$$
h_q(p,c)=\max\limits_{0\leq t\leq 1} |H_q(t)|\ \ \ (0\leq q< p)\ \ \ \text{and}\ \ \ h_p(p,c)=|H_p(0)|.
$$
Now we investigate the error functions $h_q(p,c)$ for $q=p, q=0, 0<q<p,$ respectively.

Suppose first that $q=p$. Then $r=0$ and $k=mp$. In this case we trivially have $h_p(p,c)=1-c$.

Assume next that $q=0$. Then $0\leq k<p$. Put
$$
t_0:=p(c\sqrt{p^2+1}-p).
$$
A simple calculation yields that $0\leq t_0\leq 1$, and that $H_0$ is monotone increasing on the interval $[0,t_0]$ and monotone decreasing on the interval $[t_0,1]$. Moreover, we have $H_0(0)=0$ and $H_0(1)=c-1$, hence $H_0(t_0)\geq 0$. Thus we have
$$
h_0(p,c)=\max(1-c,H_0(t_0))=\max(1-c,
\sqrt{1+c^2+p^2+c^2p^2-2cp\sqrt{p^2+1}}-1).
$$

Finally, suppose that $0<q<p$, that is $p\leq k<mp$. Put
$$
t_q:={\frac{\sqrt{p^2+q^2}(\sqrt{(p^2+q^2)(p^2+(q+1)^2)}-p^2-q^2-q)}
{(q+1)\sqrt{p^2+q^2}-q\sqrt{p^2+(q+1)^2}}}.
$$
A simple calculation gives that $0\leq t_q\leq 1$, and that $H_q$ is monotone increasing on the interval $[0,t_q]$, while monotone decreasing on the interval $[t_q,1]$. We also have
$H_q(0)=H_q(1)=c-1$. Hence $H_q(t_q)<0$ implies $-H_q(t_q)\leq 1-c$. Thus we get
$$
h_q(p,c)=\max(1-c, H_q(t_q))=
\max\left(1-c,
c\sqrt{\frac{2}{1+\sqrt{1-{\frac{p^2}{(p^2+q^2)(p^2+(q+1)^2)}}}}}-1\right).
$$

Now we calculate the error function
\[
h(p,c):=\limsup_{p\mid n,\ n\geq k\geq 0}
\left|\frac{{\mathcal{W}}(n,k)}{\sqrt{n^2+k^2}}-1\right|=
\max\limits_{0\leq q\leq p}h_q(p,c).
\]
Observe first that for fixed $p$ and $c$ the function $h_q(p,c)$ is monotone decreasing in $q$ with $2\leq q\leq p$. Hence $h_q(p,c)\leq h_1(p,c)$ for $q=2,\hdots,p$. Further, again by Maple, we obtain that for any $c$ with
${\frac{p}{\sqrt{p^2+1}}}<c\leq 1$
$$
c\sqrt{\frac{2}{1+\sqrt{1-{\frac{p^2}{(p^2+1)(p^2+4)}}}}}\leq
\sqrt{1+c^2+p^2+c^2p^2-2cp\sqrt{p^2+1}}
$$
holds, which implies $h_1(p,c)\leq h_0(p,c)$. Hence
$$
h(p,c)=\max(1-c,\sqrt{1+c^2+p^2+c^2p^2-2cp\sqrt{p^2+1}}-1)
$$
and the theorem follows.
\qed
\end{pf}

The following corollaries provide the m.r.errors ${\mathcal E}_p^B$ (when
$c=c_p^B$) and ${\mathcal E}_p^D$ (when $c=1$), respectively.

\begin{cor}
\label{c=cpb}
Let $p$ be a positive integer. Then we have
$$
c_p^B={\frac{p\sqrt{p^2+1}+2\sqrt{p^2+1-p\sqrt{p^2+1}}-2}{p^2}}.
$$
That is, the sequence $\mathcal{A}=\mathcal{A}_{c^B_p}$ of period $p$ given by
$\mathcal{A}=\overline{{\mathcal N}_{c_p^B}}$ yields the smallest m.r.error
among all sequences $\mathcal{A}_c$ of period $p$. Moreover, the error is given
by
$$
\mathcal{E}_p^B=1-c^B_p=
{\frac{p^2+2-p\sqrt{p^2+1}-2\sqrt{p^2+1-p\sqrt{p^2+1}}}{p^2}}
$$
$$
={\frac{1.5-\sqrt{2}}{p^2}}+O\left( {\frac{1}{p^4}}\right)
\approx {\frac{0.0858}{p^2}}+O\left( {\frac{1}{p^4}}\right).
$$
\end{cor}

\begin{pf}
Put
$$f(c)=1-c \mbox{   and   } g(c)=\sqrt{1+c^2+p^2+c^2p^2-2cp\sqrt{p^2+1}}-1.
$$
A straightforward computation shows that $f$ is strictly monotone
decreasing, while $g$ is strictly monotone increasing for
${\frac{p}{\sqrt{p^2+1}}}<c\leq 1$. Hence there is a unique solution of the equation $f(c)=g(c)$ in this interval. By Theorem \ref{aaerror} this solution is given by
$$
c_p^B={\frac{p\sqrt{p^2+1}+2\sqrt{p^2+1-p\sqrt{p^2+1}}-2}{p^2}}.
$$
Thus the statement follows.
\qed
\end{pf}

\begin{cor}
\label{c=1}
Let $p$ be a positive integer. Then the sequence $\mathcal{A}=\mathcal{A}_{1}$
of period $p$ given by $\mathcal{A}=\overline{{\mathcal N}_{1}}$ (corresponding
to the choice $c=1$) has m.r.error
$$
\mathcal{E}_p^D=\sqrt{(\sqrt{p^2+1}-p)^2+1}-1={\frac{1}{8p^2}}
+O\left({\frac{1}{p^4}}\right)
={\frac{0.125}{p^2}}+O\left( {\frac{1}{p^4}}\right).
$$
\end{cor}

\begin{pf}
On substituting $c=1$ into the formula of Theorem \ref{aaerror}, the statement
follows immediately.
\qed
\end{pf}

Now we give the best approximating sequences realizing the minimal maximum relative
error for $5\times 5$ matrices ($p=2$) in Theorem \ref{borge2} and for $7\times 7$ matrices
($p=3$) in Theorem \ref{borge3}, respectively.

\begin{thm}
\label{borge2}
Let $\frac{2}{\sqrt{5}}<c\leq 1.$ Let $\mathcal{A}_c=\overline{\mathcal{N}_c}$
be the corresponding sequence on $M_2$. Then the minimal m.r.error to the
Euclidean distance among the neighborhood sequences $\mathcal{A}_c$ is attained
if and only if
\[
\ \ \ c=c_2^B,\ \ \ W_1=s \ \ \ \mbox{and}\ \ \ u \leq W_2 \leq v,
\]
where
\[
s=\frac{5-\sqrt{5}+\sqrt{25-10\sqrt{5}}}{2} \approx 2.1943,
\]
\[
u=\frac{2\sqrt{2}}{\sqrt{5}}s \approx 2.7756
\ \ \ \mbox{and}\ \ \ v=2+\frac{2s}{5} \approx 2.8777.
\]
Further, the m.r.error is given by
\[
\mathcal{E}_2^B= 1-c_2^B =1-\frac{s}{\sqrt{5}}
={\frac{3-\sqrt{5}-\sqrt{5-2\sqrt{5}}}{2}} \approx 0.0187.
\]
\end{thm}

\begin{pf}
For any even $n$ with $0 \leq k \leq n$ the possible steps are $(2,0)$ of
length $2$, $(2,1)$ and $(2,-1)$ of length $W_1$, and $(2,2)$ and $(2,-2)$ of
length $W_2$. From Lemma \ref{twostep} and the inequality $2<W_1<W_2$ we see
that the path from $(0,0)$ to $(n,k)$ of minimal length consists of $k$ steps
$(2,1)$ and $\frac{n}{2}-k$ steps $(2,0)$ if $0 \leq k \leq n/2$ and of $k-n/2$
steps $(2,2)$ and $n-k$ steps $(2,1)$ if $\frac{n}{2} \leq k \leq n$. Hence we
have for the induced length function
\[
\mathcal{W}(n,k)=
\left\{
\begin{array}{ll}
kW_1+n-2k,&\mbox{if}\ k \leq \frac{n}{2},\\
(n-k)W_1+(k-\frac{n}{2} )W_2,&\mbox{otherwise}.
\end{array}
\right.
\]
Put $t=k/n$. Then the error function is given by
\[
h(W_1,W_2):=\limsup_{2\mid n,\ n\geq k\geq 0}
\left|\frac{{\mathcal{W}}(n,k)}{\sqrt{n^2+k^2}}-1\right|=
\]
\[
\max\left(
\max\limits_{0\leq t\leq {\frac{1}{2}}}
\left|{\frac{t(W_1-2)+1}{\sqrt{1+t^2}}}-1\right|,
\max\limits_{{\frac{1}{2}}\leq t\leq 1}
\left|{\frac{(1-t)W_1+(t-{\frac{1}{2}})W_2}
{\sqrt{1+t^2}}}-1\right|\right).
\]
Our aim is to choose $W_1$ and $W_2$ such that $h(W_1,W_2)$ is minimal. For
fixed $W_1$, define the function $H_0:\ \R_{\geq 0}\to\R$ by
\[
H_0(t)={\frac{t(W_1-2)+1}{\sqrt{1+t^2}}}.
\]
Put $t_0=W_1-2$. We observe that $H_0$ is monotone increasing on
$[0,t_0]$ and monotone decreasing on $[t_0,\infty)$. Hence, as $H_0(0)=1$,
\[
\max\limits_{0\leq t\leq {\frac{1}{2}}} (|H_0(t)-1|)=
\max\left(H_0(t_0)-1,1-H_0\left({\frac{1}{2}}\right)\right)
\]
\[
=\max\left(\sqrt{W_1^2-4W_1+5}-1,1-{\frac{W_1}{\sqrt{5}}}\right)
\]
if $W_1\leq 5/2$ and
\[
\max\limits_{0\leq t\leq {\frac{1}{2}}}
(|H_0(t)-1|)=H_0\left({\frac{1}{2}}\right)-1=
{\frac{W_1}{\sqrt{5}}}-1
\]
otherwise. Clearly,
\begin{equation}
\label{tag1}
\min_{W_1,W_2} (h(W_1,W_2))\geq
\min_{W_1} \max\left(\sqrt{W_1^2-4W_1+5}-1,
\left|1-{\frac{W_1}{\sqrt{5}}}\right|\right).
\end{equation}
A calculation gives that the minimum of the right-hand side is achieved
for
\[
W_1=s:={\frac{5-\sqrt{5}+\sqrt{25-10\sqrt{5}}}{2}}\approx 2.1943
\]
and equals
\[\sqrt{s^2-4s+5}-1=1-{\frac{s}{\sqrt{5}}}
={\frac{3-\sqrt{5}-\sqrt{5-2\sqrt{5}}}{2}}\approx 0.0187.
\]
Now we fix the value $s$ of $W_1$, and show that we can choose $W_2$ in a
way to have equality in (\ref{tag1}). In fact we completely describe the
set of the appropriate $W_2$-s. Consider the maximum over $t\in [1/2,1]$.
For fixed $W_2$, define the function $H_1:\ \R_{\geq 0} \to\R$ by
\[
H_1(t)=
{\frac{(1-t)W_1+(t-{\frac{1}{2}})W_2}{\sqrt{1+t^2}}}.
\]
Observe that $H_1$ attains its maximum at
$t_1:={\frac{2(W_2-W_1)}{2W_1-W_2}}$ (which is positive) and further,
$H_1$ is monotone increasing in $[0,t_1]$ and monotone decreasing in
$[t_1,\infty)$. Hence
\[
\max\limits_{{\frac{1}{2}}\leq t\leq 1} (|H_1(t)-1|)=
\max\left(1-H_1\left({\frac{1}{2}}\right),H_1(t_1)-1,1-H_1(1)\right)=
\]
\[
=\max\left(1-{\frac{W_1}{\sqrt{5}}},
{\frac{\sqrt{(2W_1-W_2)^2+4(W_2-W_1)^2}}{2}}-1,
1-{\frac{W_2}{2\sqrt{2}}}\right)
\]
if $1/2\leq t_1\leq 1$, and
\[
\max\limits_{{\frac{1}{2}}\leq t\leq 1} (|H_1(t)-1|)=
\max\left(|H_1\left({\frac{1}{2}}\right)-1|,|H_1(1)-1|\right)=
\]
\[
=\max\left(1-{\frac{W_1}{\sqrt{5}}},
\left|{\frac{W_2}{2\sqrt{2}}}-1\right|\right)
\]
otherwise. By our choice of $W_1$, we have that
\[
1-{\frac{W_1}{\sqrt{5}}}=
\left|1-{\frac{s}{\sqrt{5}}}\right| \approx 0.0187.
\]
The values of $\left|{\frac{W_2}{2\sqrt{2}}}-1\right|$ and
$\left|{\frac{\sqrt{(2W_1-W_2)^2+4(W_2-W_1)^2}}{2}}-1\right|$
do not exceed this value if and only if $u \leq W_2 \leq v$ where $u$
and $v$ are defined in the statement of the theorem. We conclude that $h(W_1,W_2)$
attains its minimum $1-{\frac{s}{\sqrt{5}}}$ if $W_1=s$ and $u\leq W_2\leq v$.

The above argument shows that $\mathcal{E}^B_2=1-\frac{W_1}{\sqrt{5}}$.
Hence the minimum among neighborhoods ${\mathcal N}_c$ is realized for
$c=c^B_2=\frac{W_1}{\sqrt{5}}$ and for no other value of $c$.
\qed
\end{pf}

\begin{thm}
\label{borge3}
Let $\frac{3}{\sqrt{10}}<c\leq 1.$ Let $\mathcal{A}_c=\overline{\mathcal{N}_c}$
be the corresponding sequence on $M_3$. Then the minimal m.r.error to the
Euclidean distance among the neighborhood sequences $\mathcal{A}_c$ is attained
if and only if
\[
 c=c^B_3, \ \ W_1=s,\ u \leq W_2 \leq v,\ q \leq W_3 \leq r,
\]
where
\[
s={\frac{30-2\sqrt{10}+2\sqrt{100-30\sqrt{10}}}{9}}\approx 3.1340,
\]
\[
u=\frac{\sqrt{13}}{\sqrt{10}}s\approx 3.5733,
\]
\[
v={\frac{143s-18\sqrt{13}+6\sqrt{1690-143\sqrt{13}s}}
{121}}\approx 3.5944,
\]
\[
q={\frac{3s}{\sqrt{5}}}\approx 4.2047,
\]
\[
r=\frac{3\sqrt{13s^2-52\sqrt{10}s+520-10W_2^2}}{13\sqrt{10}}+\frac{15W_2}{13},
\]
and in the definition of $r$, $W_2$ can be any number with $u\leq W_2\leq v$.
Further, the m.r.error is given by
\[
\mathcal{E}_3^B=1-c^B_3=1-{\frac{s}{\sqrt{10}}}
={\frac{11-3\sqrt{10}-2\sqrt{10-3\sqrt{10}}}{9}}\approx 0.0089.
\]
\end{thm}

\begin{pf}
Let $3|n$ and $0\leq k\leq n.$ The possible steps are $(3,0)$ of length $3$,
$(3,\pm 1)$ of length $W_1$, $(3,\pm 2)$ of length $W_2$, and $(3,\pm 3)$ of
length $W_3$. From the inequalities $3<W_1<W_2<W_3$ it follows that the path
from $(0,0)$ to $(n,k)$ of minimal length consists of $k$ steps $(3,1)$ and
$\frac{n}{3}-k $ steps $(3,0)$ if $0\leq k\leq\frac{n}{3}$; of $k-\frac{n}{3}$
steps $(3,2)$ and $\frac{2n}{3}-k$ steps $(3,1)$ if
$\frac{n}{3} \leq k \leq \frac{2n}{3}$; of $k-\frac{2n}{3}$ steps $(3,3)$ and
$n-k$ steps $(3,2)$ if $ \frac{2n}{3} \leq k \leq n$. Hence we have for the
induced length function
\[
\mathcal{W}(n,k)=
\left\{
\begin{array}{ll}
kW_1+n-3k,&\mbox{if}\ k\leq n/3,\\
(2n/3-k)W_1+(k-n/3)W_2,&\mbox{if}\ n/3 < k \leq 2n/3,\\
(n-k)W_2+(k-2n/3)W_3,&\mbox{otherwise}.
\end{array}
\right.
\]
Put $t=k/n$, and define the functions
$H_i:\ \R_{\geq 0}\to\R$ $(i=0,1,2)$ by
\[
H_0(t)={\frac{t(W_1-3)+1}{\sqrt{1+t^2}}},\ \ \ \ \
H_1(t)={\frac{\left({\frac{2}{3}}-t\right)W_1+\left(t-{\frac{1}{3}}\right)W_2}
{\sqrt{1+t^2}}}
\]
and
\[
H_2(t)={\frac{(1-t)W_2+\left(t-{\frac{2}{3}}\right)W_3}{\sqrt{1+t^2}}}.
\]
Then for fixed $W_1,W_2,W_3$ the error of approximation is given by
\[
h(W_1,W_2,W_3)=
\max\left(\max\limits_{0\leq t\leq {\frac{1}{3}}}|H_0(t)-1|,
\max\limits_{{\frac{1}{3}}\leq t\leq {\frac{2}{3}}}|H_1(t)-1|,
\max\limits_{{\frac{2}{3}}\leq t\leq 1}|H_2(t)-1|\right).
\]
Let
\[
t_0=W_1-3,\ t_1={\frac{3(W_2-W_1)}{2W_1-W_2}},\
t_2={\frac{3(W_3-W_2)}{3W_2-2W_3}},
\]
and observe that all $t_0$, $t_1$ and $t_2$ are positive. By
differentiation and following standard calculus, we get that for $i=0,1,2$,
$H_i$ is monotone decreasing if $t_i\not\in [i/3,(i+1)/3]$, and that $H_i$
is monotone increasing in $[i/3,t_i]$ and monotone decreasing in
$[t_i,(i+1)/3]$ otherwise. Hence from $H_0(0)=1$ we get that
\[
\max\limits_{0\leq t\leq {\frac{1}{3}}} (|H_0(t)-1|)=
\max\left(H_0(t_0)-1,\left|H_0\left({\frac{1}{3}}\right)-1\right|\right)=
\]
\[
=\max\left(\sqrt{W_1^2-6W_1+10}-1,
\left|{\frac{W_1}{\sqrt{10}}}-1\right|\right).
\]
Hence obviously,
\begin{equation}
\label{tag3}
\min_{W_1,W_2,W_3} h(W_1,W_2,W_3)\geq
\min_{W_1} \max\left(\sqrt{W_1^2-6W_1+10}-1,
\left|1-{\frac{W_1}{\sqrt{10}}}\right|\right).
\end{equation}

By a simple calculation we get that the minimum of the right-hand side is
achieved for
\[
W_1=s:={\frac{30-2\sqrt{10}+2\sqrt{100-30\sqrt{10}}}{9}}\approx 3.1340
\]
and equals
\[
M:=\sqrt{s^2-6s+10}-1=1-{\frac{s}{\sqrt{10}}}
={\frac{3\sqrt{10}+7+2\sqrt{10-3\sqrt{10}}}{9}}\approx 0.0089.
\]
Now we fix the value $s$ of $W_1$, and show that we can choose $W_2$ and $W_3$
in a way to have equality in (\ref{tag3}). More precisely, we completely
describe the set of the appropriate pairs $(W_2,W_3)$. For this purpose, first
we consider the maximum of $H_1$ over $t\in [1/3,2/3]$. In a similar manner as
in the proof of Theorem \ref{borge2}, we obtain that
\[
\max\limits_{{\frac{1}{3}}\leq t\leq {\frac{2}{3}}} (|H_1(t)-1|)=
\max\left(\left|H_1\left({\frac{1}{3}}\right)-1\right|,H_1(t_1)-1,
\left|H_1\left({\frac{2}{3}}\right)-1\right|\right)=
\]
\[
=\max\left(\left|{\frac{W_1}{\sqrt{10}}}-1\right|,
{\frac{\sqrt{{(2W_1-W_2)}^2+9{(W_2-W_1)}^2}}
{3}}-1,\left|{\frac{W_2}{\sqrt{13}}}-1\right|\right).
\]
Using our choice for $W_1$, a simple calculation gives that the above maximum
does not exceed the value of $M$ precisely when $u\leq W_2\leq v$, where $u$
and $v$ are defined in the statement of the theorem. So let $W_2$ be any fixed
number from the interval $[u,v]$, and consider the the maximum of $H_2$ over
$t\in [2/3,1]$. Now we get that
\[
\max\limits_{{\frac{2}{3}}\leq t\leq 1} (|H_2(t)-1|)=
\max\left(\left|H_2\left({\frac{2}{3}}\right)-1\right|,
H_2(t_2)-1,|H_2(1)-1|\right)=
\]
\[
=\max\Big(\left|{\frac{W_2}{\sqrt{13}}}-1\right|,
{\frac{\sqrt{{(3W_2-2W_3)}^2+9{(W_3-W_2)}^2}}{3}}-1,
\left|{\frac{W_3}{3\sqrt{2}}}-1\right|\Big).
\]
Using our choice for $W_1$ and $W_2$, a simple calculation yields that the
above maximum is not larger than $M$ if and only if $q\leq W_3\leq r$, where
$q$ and $r$ are given in the statement. (Note that $4.2766<r<4.2804$.)

The above argument shows that $\mathcal{E}^B_3=1-\frac{W_1}{\sqrt{10}}$.
Hence the minimum among neighborhoods ${\mathcal N}_c$ is realized for
$c=c^B_3=\frac{W_1}{\sqrt{10}}$, and the theorem follows.
\qed
\end{pf}

\section{Equivalence of m.r.errors for $M_p$ neighborhoods}
\label{eqapperr}

In this section we compute the m.r.errors $E^B_p$, $E^C_p$ and $E^D_p$.
First we introduce neighborhoods $N_c$ on $M_p$ defined by
$N_c(0,0)=\infty, N_c(n,0) = N_c(0,n)=|n|$ for $0<|n|\leq p$,
$N_c(n,k) = c \sqrt{n^2+k^2}$ for $(n,k) \in M_p, nk \neq 0$. Let $W_c$
denote the length function induced by the sequence $\overline {N_c}$.
We show that the corresponding m.r.error $E_c$ satisfies $E_c= \mathcal{E}_c$
for every considered value of $c$. It then follows that
$E^B_p=\mathcal{E}_p^B$ and $E^D_p = \mathcal{E}_p^D$ for every $p \geq 1$.

\begin{lem} \label{lemN}
Let $\frac{p}{\sqrt{p^2+1}}<c\leq 1$. There is a shortest $\overline{N_c}$-path from $(0,0)$ to $(mp,k)$ with $0 \leq k \leq m$ which consists of steps of the form $(p,0)$ and $(p,1)$.
\end{lem}

\begin{pf} Suppose a shortest path from $(0,0)$ to $(mp,k)$ contains a step $(g,h)$ with $h<0$. Then it also contains a step $(i,j)$ with $j \geq 1$. But it is shorter to replace both steps with steps $(g,h+1)$ and $(i,j-1)$. A similar argument can be used to exclude steps $(g,h)$ with $h > 1$. So every shortest path from $(0,0)$ to $(mp,k)$ contains only steps of the forms $(g,0)$ and $(g,1)$.

If $k=m$, then taking only steps $(p,1)$ gives the shortest path length because of the triangle inequality for the Euclidean distance and the inequality $c \leq 1$. Suppose that there is a step $(g,1)$ with $g<p$ in a shortest path from $(0,0)$ to $(mp,k)$ with $0 \leq k <m$. Then there is also a step $(h,0)$ with $h>0$. But we can replace both steps with steps $(g+1,1)$ and $(h-1,0)$ and make the path shorter.
Therefore all the steps of the form $(g,1)$ are of the form $(p,1)$. The remaining steps can be combined to steps of the form $(p,0)$.
\qed
\end{pf}

\begin{lem} \label{corN}
Let $p$ be fixed. Let $\frac{p}{\sqrt{p^2+1}} < c \leq 1$.
The m.r.error of the neighborhood sequence $\overline{N_c}$ is equal to $\mathcal{E}^D_p$ if $c=1$ and equal to $\mathcal{E}_p^B$ if $c$ assumes the value $c^B_p$ from Corollary \ref{c=cpb}.
\end{lem}

\begin{pf} Because of symmetry it suffices only to consider points $(n,k)$ with $0 \leq k \leq n$.
First let $c=1$. By definition $N(n,k) = \sqrt{n^2+k^2}$ for $(n,k) \in M_p$. Hence the induced length function satisfies $W_1(n,k)\geq|(n,k)|$ for all $(n,k)\in\Z^2$. Thus
$$
\min ~{\frac{W_1(n,k)}{\sqrt{n^2+k^2}}} \geq 1
$$
where the minimum is taken over all $(n,k)\in \Z^2$ with $(n,k) \neq (0,0)$. On the other hand, by Lemma \ref{lemN}, the shortest $\overline{N_1}$ path from $(0,0)$ to $(mp,k)$ with $0 \leq k \leq m$ consists of steps of the forms $(p,0)$ and $(p,1)$ which have lengths $p$ and $\sqrt{p^2+1}$, respectively. Hence $W_1(mp,k) = \mathcal{W}_1(mp,k)$ for $0 \leq k \leq m$. If $n=mp+r$ with $0 \leq r<p$, then $|W_1(n,k) - W_1(mp,k)| <p$. Note that in view the proof of Theorem \ref{aaerror} (in particular, since $h_0(p,c)\geq h_i(p,c)$ for all $i\geq 1$ there) we have
$$
{\underset{0\leq k\leq mp}{\limsup_{|(mp,k)|\to\infty}}} {\frac{\mathcal{W}_1(mp,k)}{|(mp,k)|}}=
{\underset{0\leq k\leq m}{\limsup_{|(mp,k)|\to\infty}}} {\frac{\mathcal{W}_1(mp,k)}{|(mp,k)|}}.
$$
Thus on the one hand it follows that
$$
{\underset{0\leq k\leq n}{\limsup_{|(n,k)|\to\infty}}} {\frac{W_1(n,k)}{|(n,k)|}}=
{\underset{0\leq k\leq mp}{\limsup_{|(mp,k)|\to\infty}}} {\frac{W_1(mp,k)}{|(mp,k)|}}\geq
$$
$$
\geq {\underset{0\leq k\leq m}{\limsup_{|(mp,k)|\to\infty}}} {\frac{W_1(mp,k)}{|(mp,k)|}}=
{\underset{0\leq k\leq m}{\limsup_{|(mp,k)|\to\infty}}} {\frac{\mathcal{W}_1(mp,k)}{|(mp,k)|}}=
{\underset{0\leq k\leq mp}{\limsup_{|(mp,k)|\to\infty}}} {\frac{\mathcal{W}_1(mp,k)}{|(mp,k)|}}.
$$
On the other hand, by $W_1(mp,k)\leq\mathcal{W}_1(mp,k)$ for all $m,p$ and $k$, we also have that
$$
{\underset{0\leq k\leq mp}{\limsup_{|(mp,k)|\to\infty}}} {\frac{\mathcal{W}_1(mp,k)}{|(mp,k)|}}\geq
{\underset{0\leq k\leq mp}{\limsup_{|(mp,k)|\to\infty}}} {\frac{W_1(mp,k)}{|(mp,k)|}}=
{\underset{0\leq k\leq n}{\limsup_{|(n,k)|\to\infty}}} {\frac{W_1(n,k)}{|(n,k)|}}.
$$
Hence
$$
{\underset{0\leq k\leq n}{\limsup_{|(n,k)|\to\infty}}} {\frac{W_1(n,k)}{|(n,k)|}}=
{\underset{0\leq k\leq mp}{\limsup_{|(mp,k)|\to\infty}}} {\frac{\mathcal{W}_1(mp,k)}{|(mp,k)|}}
$$
and by
$$
\limsup_{|(mp,k)|\to\infty} {\frac{\mathcal{W}_1(mp,k)}{|(mp,k)|}}=
1+\mathcal{E}_p^D,
$$
the m.r.error of $\overline{N_1}$ equals $\mathcal{E}_p^D$.

Next let $c= c^B_p = 1 - \mathcal{E}_p^B$. Then $\frac{p}{\sqrt{p^2+1}} <c<1$, and,
by construction, $\mathcal{W}_c(p,0) = p, \mathcal{W}_c(p,k) = c \sqrt{p^2+k^2}$ for $0<k \leq p$,
and $W_c(n,k) = c \sqrt{n^2+k^2}$ for $0 < k \leq n \leq p.$ Hence
$$
\min\limits_{(n,k) \in M_p^* } {\frac{W_c(n,k)}{\sqrt{n^2+k^2}}} = c =
1-\mathcal{E}^B_p.
$$
Thus
$$
\liminf_{|(n,k)|\to\infty}{\frac{W_{c^B_p}(n,k)}{|(n,k)|}}=1-\mathcal{E}^B_p.
$$
On the other hand, by Lemma \ref{lemN}, the shortest $\overline{N_c}$ path
from $(0,0)$ to $(mp,k)$ with $0 \leq k \leq m$ consists of steps of the form
$(p,0)$ and $(p,1)$. By a similar reasoning as above we obtain that
$$
\limsup_{|(n,k)|\to\infty}{\frac{W_{c^B_p}(n,k)}{|(n,k)|}}=1+\mathcal{E}^B_p.
$$
Thus the  m.r.error of $\overline{N_c}$ equals $\mathcal{E}^B_p$.
\qed
\end{pf}

\begin{thm}
\label{nine}
For every $p$ we have $E_p^B=\mathcal{E}_p^B$ and $E_p^D=\mathcal{E}_p^D$.
\end{thm}

\begin{pf}
We first consider the $D$-case. Suppose the neighborhood $N$ on $M_p$ induces
a length function $W:\Z^2\to\R_{\geq 0}$ such that $W(\vec{v})\geq|\vec{v}|$
for all $\vec{v} \in \Z^2$ and $W$ has m.r.error $E^D_p$. It can only improve
the m.r.error if we replace the value $N(n,k)$ for some $(n,k) \in M_p^*$ with
a smaller value $\geq |(n,k)|$. Therefore we may assume without loss of
generality that $N=N_1$. Hence $E^D_p = \mathcal{E}^D_p$.

Now we turn to the $B$-case. Suppose a neighborhood $N$ on $M_p$ induces a
length function $W$ such that
$W(n,0) = W(0,n) = |n|$ for $n \in \Z$ and
$(1 - E^B_p)|\vec{v}| \leq W(\vec{v}) \leq (1 + E^B_p) |\vec{v}|$ for all
$\vec{v} \in \Z^2$. Without loss of generality we may replace all values
$N(n,k)$ for $(n,k) \in M_p^*$ with $|n|$ if $k=0$, with $|k|$ if $n=0$, and
with $(1-E^B_p)|(n,k)|$ otherwise. Thus $E^B_p$ equals the m.r.error of the
neighborhood sequence $\overline{N_{1-E^B_p}}$. We know from Lemma \ref{corN}
and Corollary \ref{c=cpb} that if $c=c^B_p$, then the m.r.error of
$\overline{N_c}$ equals $\mathcal{E}^B_p = 1 - c^B_p$. Hence
$E^B_p\leq\mathcal{E}^B_p$. From $N(n,k)\geq(1-E^B_p)|(n,k)|\geq c^B_p|(n,k)|$
for all $(n,k) \in M_p^*$ we obtain $W(\vec{v}) \geq W_{c^B_p}(\vec{v})$ for
all $\vec{v} \in \Z^2$. Hence
$$
1+E^B_p=\inf\limits_{N} \limsup\limits_{|\vec{v}|\to\infty}
{\frac{W(\vec{v})}{|\vec{v}|}} \geq
\limsup\limits_{|\vec{v}|\to\infty}
{\frac{W_{c^B_p}(\vec{v})}{|\vec{v}|}}= 1 + \mathcal{E}^B_p$$
by Lemma \ref{corN}. Thus $E^B_p = \mathcal{E}^B_p$.
\qed
\end{pf}

Finally, we compute the minimal m.r.error $E_p^C$ for the class of arbitrary
neighborhoods $N$ defined on $M_p$.
Observe that the m.r.error $E_p^C$ is attained by the length function $W$
corresponding to the neighborhood $N$ defined by
$w(\vec{v})=(1-E_p^C)|\vec{v}|$ for $\vec{v}\in M_p^*$, since
$\frac{N(\vec{v})}{|\vec{v}|}$ should not assume a smaller value than
$1 - E^C_p$ and the limsup-value cannot increase if we decrease some
$w(\vec{v})$. Clearly, the length function $W$ corresponding to $\overline{N}$ is just
${\frac{1}{1-E_p^C}}W_1$ where $W_1$ is the length function on $\overline{N_1}$.
Recall that $\overline{N_1}$ has m.r.error $E_p^D$. Therefore we have
\begin{equation}
\label{EE}
1+E_p^C=\limsup\limits_{|\vec{v}|\to\infty}
{\frac{W(\vec{v})}{|\vec{v}|}}=(1+E_p^D)(1-E_p^C).
\end{equation}
By a simple calculation we get $E_p^C={\frac{E_p^D}{2+E_p^D}}$. So we have proved

\begin{thm}
\label{ten}
For every $p\geq 1$ we have
$$
E_p^C={\frac{E_p^D}{2+E_p^D}}={\frac{\sqrt{2p^2+2-2p\sqrt{p^2+1}}-1}
{\sqrt{2p^2+2-2p\sqrt{p^2+1}}+1}}
={\frac{1}{16p^2}}+O\left({\frac{1}{p^4}}\right).
$$
\end{thm}

\begin{rem} Observe that $E_p^B$ is about $37\%$ larger than $E_p^C$.
This is the price to be paid for the restriction $W(n,0) = |n|$
for $n \in \Z$.
The value of $E_p^D$ is about twice the error $E_p^C$. This is due to the fact that the negative and positive
deviations in $E_p^C$ are added to the positive deviation in $E_p^D$.
\end{rem}

\section{Conclusion}

In this paper, we have determined the smallest possible maximum relative
error of chamfer distances with respect to the Euclidean distance under various conditions.
We have dealt with approximating distances from three main aspects: supposing that a horizontal/vertical step has a
weight $1$ in the local chamfer neighborhoods, majorating the Euclidean
distance, and also without any constraint. We have calculated optimal weights for small ($5\times 5$ and $7\times 7$)
neighborhoods in a certain case, as well. Our framework is embedded in the theory of neighborhood sequences with possible
generalizations in this field.

\section*{Acknowledgement}

The authors are grateful to the reviewers for their valuable comments to improve the content of the paper.
Research of the Hungarian authors was supported in part by the OTKA grants F043090, T042985, T048791, K67580, K75566, by the J\'anos Bolyai Research
Fellowship of the Hungarian Academy of Sciences, by the TECH08-2 project DRSCREEN - Developing a computer based image
processing system for diabetic retinopathy screening of the National Office for Research and Technology of Hungary
(contract no.: OM-00194/2008, OM-00195/2008, OM-00196/2008), and by the T\'AMOP 4.2.1./B-09/1/KONV-2010-0007 project, which is
implemented through the New Hungary Development Plan, cofinanced by the European Social Fund and the European Regional Development
Fund.

\end{document}